\documentstyle[twoside,fleqn,espcrc2]{article}

\newcommand{\Tr}{\mathop{\rm Tr}}

\title{\bf Noncompact Lattice Simulations of SU(2) Gauge Theory}
\author{Kevin Cahill \\Department of Physics and Astronomy,
University of New Mexico\\ Albuquerque, New Mexico 87131--1156, USA}

\begin{document}
\begin{abstract}
Wilson loops have been measured
at strong coupling, $\beta=0.5$, on a $12^4$ lattice
in noncompact simulations of pure SU(2) without gauge fixing.
There is no sign of quark confinement.
\end{abstract}

\maketitle

\section*{INTRODUCTION}
In 1980 Creutz~\cite{Creutz 80} displayed quark
confinement at moderate coupling
in lattice simulations~\cite{Wilson 74}
of both abelian and nonabelian gauge theories.
Whether nonabelian confinement
is as much an artifact of Wilson's action
as is abelian confinement remains unclear.
\par
The basic variables of
Wilson's formulation
are elements of a compact group
and enter the action only through
traces of their products.
Wilson's action has extra minima~\cite{Cahill 88}.
Mack and Pietarinen~\cite{Mack}
and Grady~\cite{Grady} have shown that these false vacua
affect the string tension.
In their simulations of SU(2),
they placed gauge-invariant
infinite potential barriers
between the true vacuum and the false vacua.
Mack and Pietarinen saw a sharp drop
in the string tension;
Grady found that it vanished.
\par
To avoid using an action that has confinement built in,
some physicists have introduced lattice actions
that are noncompact discretizations
of the continuum action
with fields as the basic variables~\cite[6--11]{Cahill 88}.
Patrascioiu, Seiler, Stamatescu,
Wolff, and Zwanziger~\cite{Patrascioiu}
performed the first noncompact simulations of SU(2)
by using simple discretizations of the classical action.
They fixed the gauge and saw a force rather like Coulomb's.
\par
It is possible to use the exact classical
action in a noncompact simulation if one
interpolates the gauge fields from their
values on the vertices of simplices~\cite[7,8]{Cahill 88} or
hypercubes.
For U(1) these noncompact formulations are
accurate for general coupling strengths~\cite{Cahill abelian}.
But in four dimensions and without gauge fixing,
such formulations are implementable only in code that is quite slow.
One may develop faster code by interpolating
across plaquettes rather than throughout simplices.
For SU(2) such noncompact simulations
agree well with perturbation theory
at very weak coupling~\cite{Cahill 89}.
\par
This report relates the results of measuring
Wilson loops at strong coupling, $\beta\equiv 4/g^2=0.5$,
on a $12^4$ lattice
in a noncompact simulation of SU(2) gauge theory
without gauge fixing or fermions.
Creutz ratios of large Wilson loops
provide a lattice estimate
of the $q\bar q$-force for heavy quarks.
There is no sign of quark confinement.
\par
\section*{THIS NONCOMPACT METHOD}
In the present simulations,
the action is free of spurious zero modes,
and it is not necessary to fix the gauge.
The fields are constant on the links
of length $a$, the lattice spacing, but are interpolated
linearly throughout the plaquettes.
In the plaquette with vertices $n$, $n+e_\mu$,
$n+e_\nu$, and $n+e_\mu+e_\nu$,
the field is
\begin{eqnarray}
A_\mu^a(x) & = & \left({x_\nu \over a} -n_\nu\right)
A_\mu^a(n+e_\nu) \nonumber \\
& & \> \mbox{} + \left(n_\nu +1-{x_\nu \over a}\right) A_\mu^a(n),
\end{eqnarray}
and the field strength is
\begin{eqnarray}
F_{\mu\nu}^a(x) & = & \partial_\nu A_\mu^a(x)-\partial_\mu A_\nu^a(x)
\nonumber \\
& & \> \mbox{} + g f_{bc}^a A_\mu^b(x) A_\nu^c(x).
\end{eqnarray}
The action $S$ is the sum over all plaquettes
of the integral over each plaquette of the
squared field strength,
\begin{equation}
S=\sum_{p_{\mu\nu}}
{a^2 \over 2} \int \! dx_\mu dx_\nu F_{\mu\nu}^c(x)^2.
\end{equation}
The mean-value in the vacuum of
a euclidean-time-ordered operator $W(A)$
is approximated by a normalized multiple integral
over the $A_\mu^a(n)$'s
\begin{equation}
\langle {\cal T} W(A) \rangle_0  \approx
{\int
e ^ {-S(A)}  W(A)
\prod_{\mu,a,n}  dA_\mu^a(n)
\over \strut
 \int
e ^ {-S(A)}
\prod_{\mu,a,n}  dA_\mu^a(n)
}
\end{equation}
which one may compute numerically~\cite{Creutz 80}.
I used {\sc macsyma} to write the {\sc fortran} code~\cite{Cahill MACSYMA}.
\par
\section*{CREUTZ RATIOS}
The quantity normally used to study confinement
in quarkless gauge theories is the Wilson loop
$W(r,t)$
which is the mean-value in the vacuum of the
path-and-time-ordered exponential
\begin{equation}
W(r,t) = {1\over d} \, \left\langle
{\cal PT} \exp \left(-i g\oint \! A_\mu^a T_a dx_\mu\right)
\right\rangle_0
\end{equation}
divided by the dimension $d$ of the matrices
$T_a$ that represent the generators of the gauge group.
Although Wilson loops vanish~\cite{Cahill and Stump}
in the exact theory,
Creutz ratios $\chi (r,t)$
of Wilson loops defined~\cite{Creutz 80}
as double differences of logarithms of Wilson loops
\begin{eqnarray}
\lefteqn{\chi (r,t) = - \log W(r,t) -\log W(r-a,t-a)}  \nonumber \\
& & \> \mbox{} + \log W(r-a,t) +\log W(r,t-a)
\end{eqnarray}
are finite. For large $t$,
the Creutz ratio $\chi (r,t)$ approximates
($a^2$ times) the force between a quark
and an antiquark separated by the distance $r$.
\par
For a compact Lie group with $N$ generators $T_a$
normalized as $\Tr (T_aT_b)=k\delta_{ab}$,
the lowest-order perturbative formula for the
Creutz ratio is
\begin{eqnarray}
\chi(r,t) & = & {N \over 2 \pi^2 \beta}
\left[ - f(r,t) - f(r-a,t-a) \right. \nonumber \\
 & & \;\qquad \; \left. + \, f(r,t-a) + f(r-a,t)\right]
\label{chipert=}
\end{eqnarray}
where the function $f(r,t)$ is
\begin{eqnarray}
f(r,t) & = & {r\over t} \arctan{r\over t}
+{t\over r} \arctan{t\over r}\nonumber \\
 & & \qquad - \, \log\left({a^2\over r^2} + {a^2\over t^2}\right)
\label{f=}
\end{eqnarray}
and $\beta$ is the inverse coupling $\beta=d/(kg^2)$.
\par
\section*{MEASUREMENTS AND RESULTS}
To measure Wilson loops and their Creutz ratios,
I used a $12^4$ periodic lattice, a heat bath,
and 20 independent runs with cold starts.
The first run began with 25,000 thermalizing
sweeps at $\beta = 2$
followed by 5000 at $\beta=0.5$;
the other nineteen runs began at $\beta = 0.5$
with 20,000 thermalizing sweeps.
In all I made 59,640 measurements, 20 sweeps apart.
I used a version Parisi's trick~\cite{Parisi}
that respects the dependencies that occur
in corners of loops and between lines
separated by a single lattice spacing.
The values of the Creutz ratios so obtained
are listed in the table
along with the tree-level theoretical values
as given by eqs.(\ref{chipert=}--\ref{f=}).
I estimated the errors
by the jackknife method~\cite{jackknife},
assuming that all measurements were independent.
Binning in small groups made little difference.

$$
\vbox{
\hbox{\sl Noncompact Creutz ratios at $\beta = 0.5$}
\vskip 1pt
\vbox{\offinterlineskip
\hrule
\halign{&\vrule#&
  \strut\quad#\hfil\quad\cr
height2pt&\omit&&\omit&&\omit&\cr
&${r\over a}\times{t\over a}$\hfil&&Monte Carlo\hfil&&Order $1/\beta$\hfil&\cr
height2pt&\omit&&\omit&&\omit&\cr \noalign{\hrule}
height2pt&\omit&&\omit&&\omit&\cr
&$2\times2$&&0.23107(4)&&0.39648&\cr
height2pt&\omit&&\omit&&\omit&\cr \noalign{\hrule}
height2pt&\omit&&\omit&&\omit&\cr
&$3\times3$&&0.03576(11)&&0.13092&\cr
height2pt&\omit&&\omit&&\omit&\cr \noalign{\hrule}
height2pt&\omit&&\omit&&\omit&\cr
&$4\times4$&&0.00485(29)&&0.06529&\cr
height2pt&\omit&&\omit&&\omit&\cr \noalign{\hrule}
height2pt&\omit&&\omit&&\omit&\cr
&$5\times5$&&0.00094(82)&&0.03913&\cr
height2pt&\omit&&\omit&&\omit&\cr \noalign{\hrule}
height2pt&\omit&&\omit&&\omit&\cr
&$6\times6$&&-0.00149(226)&&0.02608&\cr
height2pt&\omit&&\omit&&\omit&\cr}
\hrule}}
$$
\par
If the static force between heavy quarks is
independent of distance,
then the Creutz ratios $\chi(r,t)$ for large $t$ should be
independent of $r$ and $t$.
The measured $\chi(r,t)$'s are smaller than
their tree-level perturbative counterparts.
The measured $\chi(r,t)$'s also
fall faster with increasing loop size.
There is no sign of confinement.
\section*{PLAUSIBLE INTERPRETATIONS}
\par
Why don't noncompact simulations display quark confinement?
Here are some possible answers:
\begin{enumerate}
\item Although noncompact methods have approximate forms of
all continuum symmetries, including gauge invariance,
they lack an exact lattice gauge invariance.
It may be possible to impose a kind of lattice gauge invariance by having
the fields randomly make suitably weighted
gauge transformations of the compact form
\begin{eqnarray}
\lefteqn{\exp[-igaA_\mu^{\prime b}(n)T_b] =} \nonumber\\
& & U(n+e_\mu)\exp[-igaA_\mu^b(n)T_b]U(n)^\dagger
\end{eqnarray}
or of some noncompact form.
\item The Maxwell-Yang-Mills action contains
squares of (covariant) curls of gauge fields,
not squares of derivatives of gauge fields.
Thus the gauge fields may vary markedly
from one link to the next.  In fact in these
noncompact simulations, the ratio of
differences of adjacent gauge fields
to their moduli
\begin{equation}
{\langle |A_\mu^b(n+e_\nu) - A_\mu^b(n)| \rangle
\over \langle |A_\mu^b (n)|\rangle}
\end{equation}
exceeds unity for all couplings
from $\beta = 0.5$ to $\beta = 60$
and all $\mu$ and $\nu$.
It is not obvious that noncompact methods
can cope with such choppiness.
\item
The noncompact lattice spacing $a_{NC}(\beta)$
is probably smaller than the compact one $a_{C}(\beta)$.
Thus noncompact methods may accommodate too small
a volume at weak coupling;
confinement might appear in noncompact
simulations done on much larger lattices or at stronger coupling.
Both possibilities would be expensive to test.
\item Perhaps SU(3), but not SU(2), confines.
\item Possibly as Gribov has suggested~\cite{Gribov},
quark confinement is due to the lightness of the up
and down quarks and not a feature of pure {\sc qcd}.
\item As Polonyi~\cite{Polonyi} has suggested,
it may be necessary in lattice definitions of
path integrals in gauge theories to enforce Gauss's law by
integrations over the group manifold weighted
by the Haar measure.
In a preliminary test of this idea,
I measured the Polyakov line of length $12a$
at $\beta = 0.5$ to be 0.00853(67) with the Haar measure
and 0.01135(1) without it.
\item
Confinement is a robust and striking phenomenon.
Maybe the true continuum theory is one
like Wilson's that can directly account for it.
The hybrid measure
\begin{equation}
e^{-\int \! \! {Dc\over kg^2} f(c,l)
\Tr(1 - \Re {\cal P} e^{ig\oint_c A_\mu^a T_a dx_\mu})} d\mu(A)
\end{equation}
reduces to Wilson's prescription if the weight functional $f(c,l)$
of the path integration over closed curves $c$
is a delta functional with support on the plaquettes
and if $d\mu(A)$ incorporates the Haar measure.
A weight functional like
$f(c,l) \sim \exp[ - (\|c\|/l)^4]$
where $\|c\|$ is the length of the curve
might give confinement for distances much longer than $l$
and perturbative {\sc qcd} for much shorter distances.
\end{enumerate}
\goodbreak
\par
\section*{ACKNOWLEDGEMENTS}
I am grateful
to H.~Barnum, M.~Creutz, G.~Kilcup, J.~Polonyi, and D.~Topa
for useful conversations and
to the Department of Energy for support under grant
DE-FG04-84ER40166.
Some of the computations reported here
were performed in collaboration with Richard Matzner
of the University of Texas at Austin.
Some were done on an RS/6000 lent by {\sc ibm},
some on Cray's at the National Energy Research
Supercomputer Center of the Department of Energy,
and some on a 710 lent by Hewlett-Packard.
Most were done on a {\sc dec}station 3100.
\goodbreak

\end{document}